\documentclass[pra,superscriptaddress,showpacs,twocolumn]{revtex4}
\usepackage{ulem}
\usepackage{amsmath}
\usepackage{amstext}
\usepackage{latexsym}
\usepackage{graphicx}
\usepackage{amsfonts}
\usepackage{epsfig}

\begin{document}
\title{Suitability of the approximate superposition of squeezed coherent states for various quantum protocols}

\author{Petr Marek}

\affiliation{School of Mathematics and Physics, The Queen's University,
  Belfast BT7 1NN, United Kingdom}

\author{M. S. Kim}

\affiliation{School of Mathematics and Physics, The Queen's University,
  Belfast BT7 1NN, United Kingdom}

\begin{abstract}
A state in a d-dimensional Hilbert space can be simulated by a state defined in a different dimension with
high fidelity. We assess how faithfully such the approximated state can perform quantum protocols, using an example of the squeezed coherent superposition state which was recently experimentally generated.
\end{abstract}
\date{\today}
\maketitle

\section{Introduction}
Quantum mechanical superposition of macroscopically distinguishable objects, also called ``cat'' states, first commented on by Schr\"oedinger in his famous gedankenexperiment \cite{gedankenexperiment}, has since sparked a great deal of interest. Significant attention has been devoted to the superposition of coherent states, which are at the border between classical and quantum states~\cite{def1,def2}. When properly generated, these states could be used in fundamental tests of quantum theory \cite{fund1,fund2,fund3,fund4}, or in various quantum information protocols, such as quantum teleportation \cite{tele1,tele2} or quantum computing \cite{qcomp}. However, the generation based on a nonlinear interaction in a Kerr medium \cite{def1}, faces a difficult obstacle in the form of accessible Kerr nonlinearities - they are much smaller than what is needed \cite{Kerr}.

Recently, different approaches have been given an attention to generate a coherent superposition state. The so-called Schr\"odinger kitten states, superposed coherent states with small amplitudes, were generated by subtracting a photon from a squeezed state \cite{kitten1,kitten2,kitten3}. Although their size limits their usability for quantum information protocols, there are ways to increase it using probabilistic linear optics scheme \cite{amplify}. Yet another scheme was successfully realized very recently~\cite{Grangier}, with various distinct features. First, the state produced is not a superposition of coherent states but squeezed coherent states.   This suggests that the state needs to be un-squeezed before it can be used for existing quantum information protocols for coherent superposition states, although we are going to show that this step needs not be necessary. Second, in contrast to earlier theoretical and experimental efforts, the state generated in \cite{Grangier} is defined in a finite Hilbert space (dimension three).

A quantum state is defined in Hilbert space of a certain dimension.  Two extreme examples are a qubit state in a two-dimensional space and a continuous-variable state in an infinite dimensional space.  In performing a quantum protocol, it is normally a tacit understanding that quantum states and operations are fixed to one particular dimension.  However, there is no reason why a quantum protocol cannot be performed using quantum systems in various dimensions.  A continuous-variable state can be projected onto a finite-dimensional space~\cite{caslav} and can be used to entangle two qubits~\cite{son}. The recent experimental realization of the superposition of squeezed coherent states~\cite{Grangier} is also an example of this.

Consider a continuous-variable state
\begin{equation}
|\Psi\rangle=c_0|0\rangle+c_1|1\rangle+c_2|2\rangle+\cdots
\label{c-state}
\end{equation}
which is to be approximated by a state in a two-dimensional Hilbert space $\{0,1\}$:
\begin{equation}
|\Phi\rangle=d_0|0\rangle+d_1|1\rangle.
\label{d-state}
\end{equation}
We take the amplitudes of both states, {\it i.e.} $c_i$ and $d_i$, real for simplicity.  In order to find how close the approximated state $|\Phi\rangle$ is to the target state $|\Psi\rangle$, we use the fidelity, which is
the overlap between the two states:
\begin{equation}
F=|\langle\Psi|\Phi\rangle|^2=(c_0d_0+c_1d_1)^2 =\left(c_0d_0+c_1\sqrt{1-d_0^2}\right)^2,
\label{fidelity}
\end{equation}
where the normalization condition has been utilized.  The fidelity is optimized to $c_0^2+c_1^2$ when $d_0/d_1=c_0/c_1$.  In other words, the sum of probabilities of the target state being in the truncated Hilbert space determines the maximal fidelity of the approximated state.  Here, we are not suggesting that this is the absolute optimum strategy to simulate $|\Psi\rangle$ by a state in a two-dimensional space.  For example, when $c_i$ and $c_j$ for arbitrary $i$ and $j$ ($i\neq j$) are larger than other amplitude factors, there is no reason why the two dimensional state cannot be defined in $|i\rangle$ and $|j\rangle$. 

When the target state is a coherent superposition state $|\Psi\rangle={\cal N}(|\alpha\rangle + |-\alpha\rangle)$, where $|\alpha\rangle$ represents a coherent state with amplitude $\alpha$.  The amplitudes of coherent states will be considered real throughout the paper. The property of the coherent state being a Poissonian-weighted sum of Fock states
gives the coefficients
\begin{equation}
c_n=\frac{\alpha^n}{\sqrt{n! \mathrm{cosh}\alpha^2}} ~~\mbox{for $n$ even}
\label{c-n}
\end{equation}
and $c_n=0$ for $n$ odd.  Throughout the paper, ${\cal N}$ denotes the normalization factor.  If the coherent superposition state is
approximated by a state in a truncated Hilbert space $\{0,2\}$, the optimum fidelity is $F=(\alpha^4+2)/(2\mathrm{cosh}\alpha^2)$.  As $\alpha$ grows, the fidelity gets smaller.  While the fidelity is $F=0.97$ for $\alpha=1$, it is already as small as $F=0.73$ for $\alpha=1.5$.

Using a superposition of zero and two photon states, Ourjoumtsev {\it et al.} approximated a squeezed coherent superposition state
\begin{equation}
|\Psi_s\rangle= \mathcal{N} S(r)(|\alpha\rangle+|-\alpha\rangle)
\label{sq-coh}
\end{equation}
where the $S(r)$ is the squeezing operator with the real parameter $r$ resulting in squeezing the $x$ quadrature variance. With use of the expansion of $|\Psi_s\rangle$ in the Fock basis, we find with the argument above that the optimum fidelity to be approximated by a state in the truncated space $\{0,2\}$ is
\begin{eqnarray}
F &=& \langle \Psi_s|0\rangle^2 + \langle \Psi_s|2\rangle^2 \nonumber \\
 &=& \frac{2\sqrt{2g}}{[1+\exp(-2\alpha^2](1+g)^5}\exp\left(\frac{-2g\alpha^2)}{1+g}\right) \nonumber \\
 &\times& \left[2(1+g)^4 + (4g\alpha^2 - 1+g^2)^2 \right],
\label{fidelity-sq}
\end{eqnarray}
where $g = \exp(-2r)$.  We can see that the fidelity can be larger for an appropriate choice of parameters.  In Ourjoumtsev {\it et al.}'s work, the fidelity they achieved is $0.99$ for $r=0.4029$ (corresponding to $3.5$ dB of squeezing) and $\alpha=\sqrt{2.6}$.  It is clear that their approximated state is nearly optimum in the truncated space $\{0,2\}$ and that the components outside of $\{0,2\}$ space are small.  However, as these components are not zero, we can ask a question if the truncation affects any of the quantum protocols when the approximate state replaces the target state.  This is the question we want to answer to in this paper, as we exemplify some protocols.

\section{Teleportation using ideal squeezed superposition state}
In order to analyze performance of the approximate superposition of squeezed coherent states for quantum protocols, we need a benchmark to compare it with. Therefore, in this section we shall study an ideal superposition state with regard to quantum teleportation protocol.
Consider a general qubit-like state in a squeezed coherent-state basis 
\begin{equation}\label{psi_sig}
|\psi_{\mathrm{sig}}\rangle = \mathcal{N}_{\mathrm{sig}}S(r)  \left(a |\alpha\rangle + b |-\alpha\rangle\right),
\end{equation}
where the normalization constant is
\begin{equation}
\mathcal{N_{\mathrm{sig}}} = \left[|a|^2+|b|^2+2 \exp(-2\alpha^2) \mbox{Re}\{a b^*\}\right]^{-1/2}.
\end{equation}
When $a=b$, $|\psi_{\mathrm{sig}}\rangle$ the same as $|\Psi_s\rangle$ in Eq.~(\ref{sq-coh}).
This state bears many similar properties to a coherent superposition state without squeezing. For both classes of states, the base states $|\alpha\rangle$ and $|-\alpha\rangle$ for a coherent superposition state and $S(r)|\alpha\rangle$ and $S(r)|-\alpha\rangle$ for its squeezed counterpart are only asymptotically orthogonal, with the overlap of $|\langle \alpha|-\alpha\rangle|^2 = \exp(-4\alpha^2)$.  However, orthogonal bases exist in the form of superpositions between either odd or even numbered Fock states: $|\pm(\alpha)\rangle \propto |\alpha\rangle \pm |-\alpha\rangle$ (normalization omitted) and $S(r)|\pm(\alpha)\rangle$. In this way, the superposition states may be treated as a qubit system rather than a continuous variable system.  These similarities suggest that it could in principle be possible to use the squeezed coherent superposition states in a same manner as the coherent superposition states, for example a quantum computation protocol, analogically to \cite{qcomp}.

In the following we shall focus on the performance of the teleportation protocol, as it is a flagship experiment based on quantum entanglement and can also be utilized for the implementation of other quantum operations \cite{Gottesman}.  For our aim to study the basic usability of the approximated squeezed coherent superposition states, firstly we need to develop the teleportation protocol for the ideal squeezed coherent superposition states.  We can then consider how faithfully the approximated state works for this protocol.

\begin{figure}
\centerline{\psfig{figure = 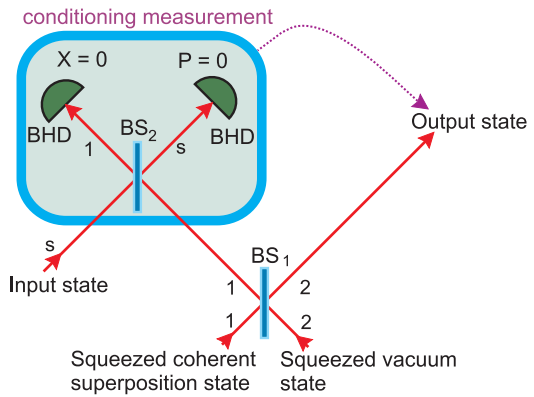,width=0.8\linewidth}}
\caption{(color online) Teleportation scheme for a squeezed coherent superposition state. BS1 and BS2 - balanced beam splitters, BHD - balanced homodyne detection. }
\label{setup}
\end{figure}

As an entangled resource to use in the setup as shown in Fig.~\ref{setup}, let us first consider a squeezed coherent superposition state in mode 1, which is mixed with a squeezed vacuum in mode 2 at a beam splitter to produce
\begin{eqnarray}\label{psi_ent}
|\psi_{\mathrm{ent}}\rangle &=& \mathcal{N}_{\mathrm{ent}} S_1(r) S_2(r)\left(|\alpha,\alpha\rangle + |-\alpha,-\alpha\rangle\right)_{1~2},\nonumber \\
\mathcal{N}_{\mathrm{ent}} &=& \left[2 + 2\exp(-|\alpha|^2/2)\right]^{-1/2}
\end{eqnarray}
where the subscripts stand for mode labels.

The first step of the teleportation protocol, as depicted in Fig.~\ref{setup} is to perform a joint measurement on modes $s$ and $1$.  The measurement outcome then tells us which unitary operation to perform on mode $2$ in order to retrieve the signal state. However, due to difficulties in the realization of joint measurements and unitary operations, it may be beneficial for the proof of principle experiment to concentrate on a single possible outcome of the joint measurement, for which no operation is required on the mode $2$. In this case, when the joint measurement heralds the desired outcome, the state of mode $2$ is accepted as a teleported one.

In order to find the required measurement explicitly, let us consider the state of the whole system after BS1,
\begin{equation}
|\psi_{\mathrm{tot}}\rangle = |\psi_{\mathrm{sig}}\rangle_s |\psi_{\mathrm{ent}}\rangle_{12},
\end{equation}
which, after mixing modes 1 and $s$ at BS2, transforms to (normalization omitted)
\begin{eqnarray}
|\psi_{\mathrm{tot}}\rangle &=& S_{\mathrm{tot}}(r) \Bigl[ a|\sqrt{2}\alpha,0,\alpha\rangle + a|0,-\sqrt{2}\alpha,-\alpha\rangle + \nonumber \\
&+& b|0,\sqrt{2}\alpha,\alpha\rangle + b|-\sqrt{2}\alpha,0,\alpha\rangle \Bigr],
\end{eqnarray}
with $S_{\mathrm{tot}}(r) = S_s(r)\otimes S_1(r)\otimes S_2(r)$.  Note that all the modes are assumed to have the same initial squeezing factor $r$.  We then use the measurement that confirms state $S(r)|0\rangle$ in mode $1$ and state $S(r)|+(\sqrt{2}\alpha)\rangle$ in mode $s$. For a coherent state superposition protocol without any squeezing, this could be done by applying photon number resolving detectors. However, with additional squeezing, a different approach has to be employed.   The squeezed vacuum is discerned by a homodyne detector,  postselecting quadrature $x_1=0$. For mode $s$, the homodyne detection of the $p_s$ quadrature, conditioned on $p_s=0$, allows to discern state $S_s(r)|+(\sqrt{2}\alpha)\rangle$ from the state $S_s(r)|-(\sqrt{2}\alpha)\rangle$, as in \cite{20}. However, although the detection of $p_s=0$ ensures that mode $s$ is not in state $S_s(r)|-(-\sqrt{2}\alpha)\rangle$, there remains a nonzero probability of it being in state $S_s(r)|0\rangle$. Similarly, the measurement outcome $x_1=0$ leaves a possibility of mode $1$ in $S_1(r)|+(\sqrt{2}\alpha)\rangle$. The output state in mode 2, conditioned on the measurement outcomes $x_1=0$ and $p_s=0$, is
\begin{eqnarray}\label{psi_out}
|\psi_{\mathrm{out}}\rangle &=&\mathcal{N}_{\mathrm{out}} \Bigl[ P_{x,0}P_{p,+} |\psi_\mathrm{sig}\rangle + P_{x,+}P_{p,0} |\psi_{\mathrm{sig}}^{S}\rangle \Bigr]
\end{eqnarray}
where
\begin{eqnarray}
\mathcal{N}_{\mathrm{out}} &=& \Bigl(|P_{p,0}|^2 |P_{x,+}|^2 + |P_{p,+}|^2|P_{x,0}|^2 +\nonumber \\
 &+& 2\mathrm{Re} [ P_{p,0}P_{x,+} P_{p,+}^* P_{x,0}^* \langle \psi_{\mathrm{sig}}|\psi_{\mathrm{sig}}^S\rangle \Bigr)^{-1/2}
\end{eqnarray}
and
\begin{eqnarray}
P_{q,i} &=& \langle i|q=0\rangle\langle q=0|j\rangle, \quad q = x,p;~~ i = +,0;\nonumber \\
|\psi_{sig}^S\rangle &=& \mathcal{N}_{sig}\left(a S_s(r)|-\alpha\rangle + b S(r)|\alpha\rangle\right).
\end{eqnarray}
The Fidelity, $F = |\langle\psi_{\mathrm{sig}}|\psi_{\mathrm{out}}\rangle|^2$, can now be evaluated with help of relations
\begin{eqnarray}\label{xprobs}
P_{p,0} &=& \frac{1}{2}P_{p,+} = \left(\frac{g}{\pi}\right)^{1/4},\nonumber \\
P_{x,0} &=& \frac{1}{2} \mbox{e}^{2\alpha^2}P_{x,+}= (\pi g)^{-1/4}, \nonumber \\
\langle \psi_{\mathrm{sig}}|\psi_{\mathrm{sig}}^S\rangle &=& \frac{(|a|^2+|b|^2)e^{-2\alpha^2} + 2\mathrm{Re}[ab^*]}{|a|^2+|b|^2+2 e^{-2\alpha^2} \mathrm{Re}[a b^*]}.
\end{eqnarray}
However, the fidelity depends on the signal state as seen in state (\ref{psi_out}) for $a = b$, in which case $|\psi_{\mathrm{sig}}\rangle = |\psi_{\mathrm{sig}}^S\rangle$ and the fidelity is consequently equal to unity.   This is why the average fidelity should be used to measure the performance of teleportation.  Alternatively, the ratio
\begin{equation}
\frac{|P_{x,0}P_{p,+}|}{|P_{x,+}P_{p,0}|} = e^{2\alpha^2}=R
\end{equation}
reliably quantifies the content of the signal in the teleported state.  If we consider states $|\psi_{\mathrm{sig}}\rangle$ and $|\psi_{\mathrm{sig}}^{S}\rangle$ to be orthogonal, we can find the lower bound of the fidelity to be $F_{min}=\frac{R^2}{1+R^2}$.  Therefore, even for attainable values of $\alpha$, for example $\alpha=\sqrt{2.6}$ in \cite{Grangier}, the teleported state is in a good agreement with the initial one.  Note, this ratio does not depend on the squeezing of the state thus the setup could also be used to teleport an un-squeezed superposition state.

A slightly different measurement needs to be used when a different entangled resource, $S_{\mathrm{tot}}(|\alpha,\alpha\rangle-|-\alpha,-\alpha\rangle)$  is employed instead of the entangled state (\ref{psi_ent}). That is because the alternative state is composed of an odd number of photons and the projection onto $S(r)|+(\alpha)\rangle$, which is used above, will lead to a state with a parity different from the initial state. In order to accommodate the difference, a projective measurement onto $S(r)|-(\sqrt{2}\alpha)\rangle$ has to be used, which can be implemented by photon number detection.  However,  a more feasible approach involving just homodyne detection can be devised \cite{20}. Let us again consider the homodyne measurement of the quadrature $p$.  For this time,  the final state is post-selected only when the measurement outcome is equal to some pre-determined value, $p = \beta$. The relevant probability amplitudes can be quickly found to be:
\begin{eqnarray}
\langle p = \beta|0\rangle &=& P_{p=\beta,0} = \left(\frac{g}{\pi}\right)^{1/4} e^{-g \beta^2/2}, \nonumber \\
\langle p = \beta|+(\alpha)\rangle &=& P_{p=\beta,+} =2  \cos 2\alpha\beta\sqrt{g}P_{p=\beta,0}, \nonumber \\
\langle p = \beta|-\rangle &=& P_{p=\beta,-} = 2i \sin 2\alpha\beta\sqrt{g}P_{p=\beta,0}.
\end{eqnarray}
For $\beta = \pi/(4\alpha \sqrt{ g})$, $P_{p=\beta,+} =0$ and the measurement approximately performs the odd-photon state projection as required.  In analogy to (\ref{psi_out}), the final state can now be expressed as
\begin{equation}
|\psi_{\mathrm{out}}'\rangle \propto P_{x,0}P_{p=\beta,-}|\psi_{\mathrm{sig}}\rangle + P_{x,+}P_{p=\beta,0}|\psi_{\mathrm{sig}}^{FS}\rangle,
\end{equation}
where
\begin{equation}
 |\psi_{\mathrm{sig}}^{FS}\rangle = \mathcal{N}_{\mathrm{sig}}S_s(r)(a|-\alpha\rangle - b|\alpha\rangle).
\end{equation}
It is again possible to assess the performance by looking at the simple ratio
\begin{equation}
 \frac{|P_{x,0}P_{p=\beta,-}|}{|P_{x,+}P_{p=\beta,0}|} = e^{2\alpha^2} \sqrt{\tanh 2\alpha^2},
\end{equation}
realizing that the portion of the teleported state is smaller than in the case of the even-number entangled state.  The difference vanishes as the amplitude of the state increases.

Note that the post-selection based on detecting a value $p = \beta$ is equivalent to displacing the initial state along the $p$ quadrature before the teleportation, and thus can implement single qubit operations such as the sign flip or phase shift~\cite{20}.

\section{Teleportation using the approximate state}
In the following, we shall consider the teleportation with the entangled resource (\ref{psi_ent}) created by  the approximate superposition state recently generated in \cite{Grangier}.  The approximate state is obtained from an initial Fock state of the excitation number $n$.  (In the actual experiment, the excitation number $n$ was $2$.  The performance of teleportation using such the approximate state is compared with that for the ideal state, after deriving an output state for a general case of an arbitrary excitation number $n$.)   The Fock state is mixed with a vacuum at a beam splitter then a state is selected in one output mode, conditioned on the homodyne measurement outcome in the other output mode.   The conditionally generated state is represented by a wave function
\begin{equation}\label{psi_cat}
\psi_{\mathrm{app}}(x_1) = x_1^n e^{-x_1^2/2} \sqrt{\frac{2^{2n}n!}{\sqrt{\pi}(2n)!}},
\end{equation}
which approximates the squeezed superposition state \cite{Grangier}.  When dealing with the approximate state, we use the formalism of wave functions as it allows for clearer explanations.    In order to create a quantum channel of the teleportation protocol in Fig.~\ref{setup}, the state is mixed at a balanced beam splitter (BS1) with the squeezed vacuum state
\begin{equation}
\psi_{\mathrm{vac}}(x_2) = e^{-x_2^2/2g} (\pi g)^{-1/4},
\end{equation}
where the squeezing parameter  $g$ is chosen to coincide with the effective squeezing parameter of the approximate state.  The wave function of the unknown signal state(\ref{psi_sig}) is
\begin{eqnarray}
\psi_{\mathrm{sig}}(x_s) &=& (\pi g)^{1/4}\mathcal{N}_{\mathrm{sig}} \times \nonumber \\
& &\left[a e^{-\frac{(x_s-\alpha\sqrt{2g})^2}{2g}} + b e^{-\frac{(x_s+\alpha\sqrt{2g})^2}{2g}}\right].
\end{eqnarray}
Conditioned on the measurement outcomes, $p_s = \beta$ and $x_1 = 0$, the wave function of the output state is
\begin{eqnarray}
\label{extra1}
\psi_{\mathrm{out}}(x) &=& \int_{-\infty}^{\infty} e^{i \beta x_s} \psi_{\mathrm{vac}}\left(\frac{x}{\sqrt{2}} - \frac{x_s}{2}\right) \nonumber \\
&\times& \psi_{\mathrm{cat}}\left(\frac{x_s}{2} + \frac{x}{\sqrt{2}}\right) \psi_{\mathrm{sig}}\left(\frac{x_s}{\sqrt{2}}\right) dx_s,
\end{eqnarray}
which, after some algebra, becomes to have the form
\begin{eqnarray}\label{psi_out_real}
\psi_{\mathrm{out}}(x) &\propto&\sum_{k=0}^n
\left(\begin{array}{c}
n \\ k
\end{array}
\right) \left(\frac{x}{\sqrt{2}}\right)^{n-k}\frac{1}{2^k} e^{-C}  \nonumber \\
&\times&\left[a e^{D\alpha}\mu_{k,+} + b e^{-D \alpha}\mu_{k,-}\right],
\end{eqnarray}
where $\mu_{k,\pm}$ denote the $k^{`\mbox{th}}$ moments of the Gaussian distribution with the mean value $\mu_{1,\pm} = B_{\pm}/A$ and variance $\mu_{2,\pm} +\mu_{1,\pm}^2 = 1/2A$ and
\begin{eqnarray}
A &=& \frac{1}{8}\left(\frac{3}{g}+1\right),\nonumber \\
B_{\pm} &=& \frac{x}{4\sqrt{2}}\left(\frac{1}{g}-1\right) \pm \frac{\alpha}{2\sqrt{g}} + i \beta, \nonumber \\
C &=& \frac{x^2}{4}\left[ 1+\frac{1}{g} -\frac{(1-1/g)^2}{8A}\right] + \alpha^2\left(1-\frac{1}{4Ag}\right) \nonumber \\
 & & + \frac{\beta^2}{A} + i\frac{\beta x}{2 \sqrt{2}A}\left(\frac{1}{g}-1\right), \nonumber \\
D &=& \frac{x}{4\sqrt{2g}A }\left(\frac{1}{g}-1 \right) + \frac{i \beta}{A}.
\end{eqnarray}
The normalization has been omitted in Eq.~(\ref{psi_out_real}).  When the initial Fock state is with an even number $n$, we do not need to perform a phase shift operation so that $\beta=0$, as explained in Section II.   A nonzero value of $\beta$ corresponds to the case when the excitation number of the initial Fock state is odd.  In this case, the phase shift value is $\beta = \pi/(4\alpha\sqrt{g})$.

\begin{figure}
\centerline{\psfig{figure = 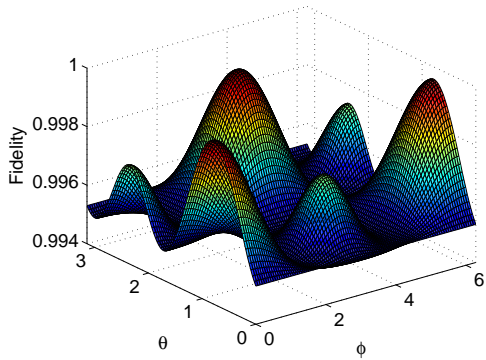,width=0.8\linewidth}}
\caption{(color online) Fidelity of teleported squeezed superposed state with $a = \cos \theta$ and $b = e^{i\phi}\sin\theta$ using ideal squeezed superposed state as a resource.  }
\label{Fid2}
\end{figure}

\begin{figure}
\centerline{\psfig{figure = 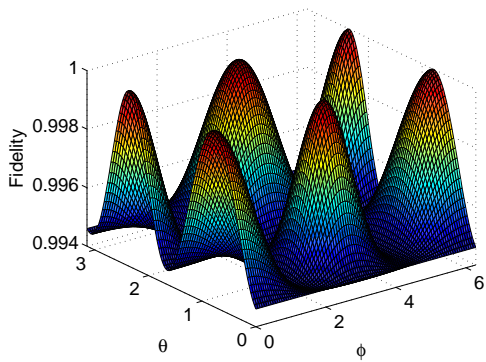,width=0.8\linewidth}}
\caption{(color online) Fidelity of teleported squeezed superposed state with $a = \cos \theta$ and $b = e^{i\phi}\sin\theta$ using approximated squeezed superposed state as a resource. }
\label{Fid1}
\end{figure}

Let us consider the actual state recently prepared in \cite{Grangier} and compare its performance with the ideal state. The output state is described by the wave function (\ref{extra1}) with $n=2$ and although it containes at most two photons it reliably approximates a squeezed coherent superposition state as analyzed earlier (See the discussions on Eq.~(\ref{fidelity-sq})) . The average fidelity is calculated as follows
\begin{equation}
F_{\mathrm{avg}} = \frac{1}{4 \pi}\int_{\Omega} F(\psi_{\mathrm{sig}}) d\Omega,
\end{equation}
where the $F(\psi_{\mathrm{sig}})$ denotes the Fidelity of teleportation for a particular input state (\ref{psi_sig}) with unique vales $a$ and $b$, {\it i.e.}
\begin{equation}
F (\psi_{\mathrm{sig}})= \left|\int_{-\infty}^{\infty} \psi_{\mathrm{sig}}^*(x)\psi_{\mathrm{out}}(x) d x\right|^2,
\end{equation}
and the integration is over the Bloch sphere of qubit-like states with basis $S(r)|\alpha\rangle$ and $S(r)|-\alpha\rangle$. The average fidelity is found to be $F_{\mathrm{avg}} = 0.9963$, for both the ideal state and the approximate state. Of course, the two states do not give the same results for any signal state. A detailed comparison can be seen in Figs.~\ref{Fid2} and \ref{Fid1}, depending on the initial state. Although the general behaviors are similar to each other, there are differences.  The maximum fidelities are obtained for the state with $a = b = 1$, where the ideal scenario gives $F = 1$ while the approximate state leads to $F = 0.9996$. The greatest difference can be observed for $a=-b=1$, where the ideal state still allows for flawless teleportation but the approximate state limits the fiedlity to $F=0.9974$.  Interestingly enough, for majority of the states the approximate superposition state leads to slightly better fidelities than the ideal one, although for all practical purposes the difference is negligible.

\section{Amplification of the superposition state}
Schr\"odinger's paradox was suggested to connect the microscopic world (atomic decay), where quantum nature is expected, to the macroscopic one (cat's destiny).  Therefore, attempts have been made to devise an amplification scheme that would allow to increase the coherent amplitude of a superposition state \cite{amplify}.  For example, for a superposition of un-squeezed coherent states, this can be achieved by mixing two identical superposition states on a balanced beam splitter, followed by a measurement on one of the modes in order to ascertain the mode being the vacuum.  Conditioned on the vacuum, an amplification is achieved in the other output mode.
The projection on the vacuum state can be implemented by an on/off detector such as an ideal avalanched photodiode, but the detection inefficiency renders this strategy experimentally not very  feasible.

\begin{figure}
\centerline{\psfig{figure = 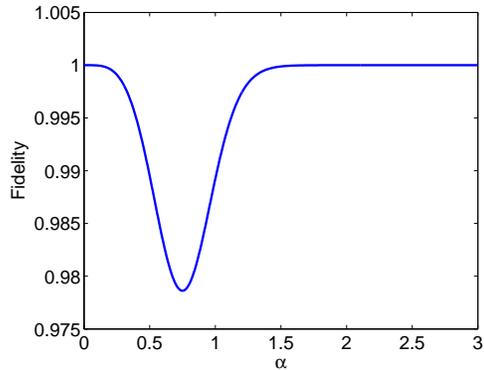,width=0.8\linewidth}}
\caption{Fidelity after the single step amplification protocol with regard to the amplitude $\alpha$ of the initial amplitude of the squeezed superposition state. The fidelity is the overlap with the squeezed superposition state of its amplitude $\alpha\sqrt{2}$.  }
\label{singleit}
\end{figure}

Let us introduce an amplification scheme for the superposition of squeezed coherent states in analogy to the amplification scheme for the coherent superposition state.  An apparent difference is that instead of the vacuum, the amplification is heralded by the measurement of the squeezed vacuum in one output mode of the beam splitter.  For this purpose, a homodyne detector may be used.  After mixing two copies of the squeezed superposition state on a balanced beam splitter, the state becomes
\begin{equation}
|\psi\rangle \propto S_{1}(r)S_2(r)\left[|0\rangle_1 |+(\sqrt{2}\alpha)\rangle_2 + |+(\sqrt{2}\alpha)\rangle_1 |0\rangle_2\right].
\end{equation}
Now, there is no need to worry about the null measurement as the amplification is conditioned on  $|x=0\rangle_2\langle x=0|$ measurement outcome.  In this case, the state transforms to
\begin{equation}\label{amplify1}
|\psi_{\mathrm{amp}}\rangle \propto S_2(r)\left[P_{x,0} |+(\sqrt{2}\alpha)\rangle_2 + P_{x,+}|0\rangle_2\right],
\end{equation}
where $P_{x,0}$ and $P_{x,+}$ are defined in Eq.~(\ref{xprobs}).  Unfortunately, this state is not ideal and after several iterations the state will be a superposition of squeezed coherent states with different amplitudes together with the squeezed vacuum. Fig.~\ref{singleit} shows the fidelity after a single step of the amplification protocol. Apparently, the non-ideal nature of the protocol is not so detrimental if the amplitude of the state is larger than unity.  Note, that the high value of fidelity for $\alpha$ close to zero is caused by the good overlap of the squeezed vacuum portion of the state (\ref{amplify1}), thus such the state is unsuitable for further amplification as numerical analysis confirms a drop of the fidelity in the repetition of the amplification protocols.  Finally, it is important to note that the actual result of amplification is independent of the squeezing thus our scheme can be used even for the amplification of a coherent superposition state without squeezing.

Let us consider the amplification of the approximate state (\ref{psi_cat}).  By mixing the two copies of the approximate states, the wave function of the total state becomes
\begin{equation}
\psi(x1,x2) = \psi_{\mathrm{app}}\left(\frac{x1-x2}{\sqrt{2}}\right) \psi_{\mathrm{app}}\left(\frac{x1+x2}{\sqrt{2}}\right).
\end{equation}
After the homodyne measurement and post-selection on $x_2 = 0$, the state transforms to
\begin{equation}
\psi_{\mathrm{amp}}(x) = \psi_{\mathrm{app}}^2\left(\frac{x}{\sqrt{2}}\right) \propto x^{2n} e^{-x^2/2}.
\end{equation}
We note that this is the same as the approximate state (\ref{psi_cat}) created from the initial $2n$-photon state, which, according to \cite{Grangier}, is an approximation of squeezed superposition state of its amplitude $\alpha = \sqrt{2n}$ with the fidelity approaching to unity as $n$ increases.  Therefore,  we can conclude that our amplification scheme based on the homodyne measurement not just increases the amplitude of the approximate state but also makes it closer to the ideal superposition state.

\section{Summary}
Quantum protocols, which employ only homodyne detecion and post-selection, can be applied equally well to both squeezed and un-squeezed superposition of coherent states. Therefore, the squeezed coherent superposition states recently generated, can be used for the proof-of-principle experiments such as quantum teleportation or single qubit gates, without any modifications. Furthermore, although the experimental realization of the squeezed superposition state is in a limited Hilbert space, it allows to teleport a signal state with about the same average fidelity as the ideal state.

In a similar vein, the amplification protocol developed for an un-squeezed coherent superposition state can be modified to accommodate a squeezed superposition state by implementing homodyne detection as the projection measurement. Although this modification leads to certain imperfections, if the initial amplitude of the superposition state is not too small, it performs rather well. When this protocol is applied to the approximate squeezed superposition state, the amplified state is again a member of the same hierarchy with the initial excitation number increased by a factor of two.  Therefore our protocol not only amplifies the state but also makes it closer to the ideal superposition state.

\medskip
This work was supported by the UK EPSRC and QIP IRC.  P. M. acknowledges support of the European Social Fund.

\end{document}